\documentclass{PoS}

\usepackage{textcomp}
\usepackage{amsfonts} 
\usepackage{amssymb} 
\usepackage{amsmath} 
\usepackage{slashed}
\usepackage{graphicx}
\usepackage{pifont}
\usepackage{multirow,lscape}
\usepackage{array}
\usepackage{longtable}
\usepackage[FIGBOTCAP,TABBOTCAP,tight]{subfigure}
\usepackage{verbatim}
\usepackage{bm}
\usepackage{comment}
\usepackage{xcolor}
\usepackage{url} 
\usepackage{epstopdf}
\usepackage{grffile}
\usepackage{multicol}

\graphicspath{{./figs/}}

\title{Quark Mass Dependence of the QCD Critical End Point in the Strong Coupling Limit}

\ShortTitle{Mass Dependence of the CEP in the Strong Coupling}

\author{\speaker{Jangho Kim}\\
        Fakult{\"a}t f{\"u}r Physik, Universit{\"a}t Bielefeld, Universit{\"a}tstasse 25, D33619 Bielefeld, Germany\\
        E-mail: \email{jangho@physik.uni-bielfeld.de}
}
\author{Wolfgang Unger\\
        Fakult{\"a}t f{\"u}r Physik, Universit{\"a}t Bielefeld, Universit{\"a}tstasse 25, D33619 Bielefeld, Germany\\
        E-mail: \email{wunger@physik.uni-bielfeld.de}
}

\abstract{
Strong coupling lattice QCD in the dual representation allows to study the full
$\mu$-$T$ phase diagram, due to the mildness of the finite density sign
problem.  
Such simulations have been performed in the chiral limit, both at finite $N_t$
and in the continuous time limit. 
Here we extend the phase diagram to finite quark masses, with an emphasis on
the low temperature first order transition.
We present our results on the quark mass dependence of the critical end point
and the first order line obtained by Monte Carlo via the worm algorithm.  
}

\FullConference{34th annual International Symposium on Lattice Field Theory\\
		24-30 July 2016\\
		University of Southampton, UK}

\begin{document}

\section{Introduction}
It is possible to investigate the full $\mu$-$T$ phase diagram using strong
coupling lattice QCD in the dual representation due to its mild sign problem.
The sign problem depends on the representation of the partition function. 
It is well known that in the strong coupling limit $\beta \equiv
\frac{2N_c}{g^2} \rightarrow 0$, i.~e.~in the absence of the gauge action, one
can make use of a dual representation due to the factorization of one-link
gauge integrals~\cite{Rossi:1984cv}.
This dual representation is well suited for Monte Carlo simulations via the
worm algorithm~\cite{Adams:2003cca,Fromm}.
Such simulations on the $\mu$-$T$ phase diagram have been carried out in the
chiral limit~\cite{deForcrand:2009dh,Unger:2011in}.
Simulation at finite quark masses have been studied in~\cite{Fromm}.
Here we extend on these studies and focus on the phase boundary for finite
quark masses. 
Also we obtain the critical end points (CEP) for finite quark masses.\\
The Lagrangian for staggered fermions $\chi$ including an anisotropy $\gamma$,
favoring temporal gauge links in order to continuously vary the
temperature, is:
\begin{align}
\label{eq:L}
\mathcal{L}_F = 
\sum_{\nu} \frac{\gamma^{\delta_{\nu 0}}}{2}\eta_{\nu}(x) 
\big( 
    e^{\mu\delta_{\nu 0}} \overline{\chi}(x) U_{\nu}(x) \chi(x+\hat{\nu}) 
  - e^{-\mu \delta_{\nu 0}} \overline{\chi}(x+\hat{\nu})U^{\dagger}_{\nu}(x) \chi(x) 
\big)\,.
\end{align}
From the Eq.~\eqref{eq:L}, one can derive the partition function in the dual
representation by integrating out the gauge links and Grassmann variables:
\begin{align}
\label{eq:dual}
Z=\sum_{\{k,n,\ell\}} 
\underbrace{\prod_{b=(x,\hat{\mu})}\frac{(N_c-k_b)!}{N_c!k_b!}\gamma^{2k_b\delta_{\hat{0},\hat{\mu}}}}_{\text{meson hoppings}}
\underbrace{\prod_{x}\frac{N_c!}{n_x!}(2am_q)^{n_x}}_{\text{chiral condensate}}
\underbrace{\prod_{\ell}w(\ell,\mu)}_{\text{baryon hoppings}}\,.
\end{align}
This partition function describes a system of mesons and baryons.
The mesons live on the bonds $b\equiv (x,\hat{\mu})$, where they hop to a
nearest neighbor $y=x+\hat{\mu}$, and the hopping multiplicity are given by
so-called dimers $k_b\in \{0,\ldots,N_c\}$.
The baryon must form self-avoiding loops.
\begin{align}
w(\ell,\mu)=\dfrac{1}{\prod_{x\in\ell}}\sigma(\ell)\gamma^{3N_{\hat{0}}} \exp{(3N_tr_{\ell}a_t\mu)}\,,
\quad \sigma(\ell)=(-1)^{r_{\ell}+N_{-}(\ell)+1}\prod_{b=(x,\hat{\mu})\in\ell}\eta_{\hat{\mu}}(x)\,,  
\end{align}
where $\ell$ denotes a baryon loop, $N_{\hat{0}}$ is the number of baryons in
temporal direction. $N_t$ is the number of lattice sites in temporal direction
and $r_{\ell}$ is the baryon winding number in temporal direction. 
$\sigma(\ell)$ is the sign.
The sign is related to staggered phase factor $\eta_{\hat{\mu}}(x)$ and the
geometry of the baryon loop $\ell$: the winding number $r_\ell$ and the number
of baryons in negative direction $N_{-}(\ell)$.
$N_{-}(\ell)$ comes from the negative sign in front of the second term in
Eq.~\eqref{eq:L}.
By the Grassmann constraint, the summation over configurations
$\sum_{\{k,n,\ell\}}$ in Eq.~\eqref{eq:L} is restricted by the following
condition.
\begin{align}
n_x + \sum_{\mu=\pm 0,\cdots,\pm d}\bigg( k_{\mu}(x)+\frac{N_c}{2}|\ell_{\mu}(x)|\bigg) = N_c
\end{align}
In the chiral condensate part, $m_q$ is the quark mass and $n_x$ is the number
of monomers at site $x$.
In the chiral limit, monomers are absent to avoid that the partition function
becomes zero. 
On the contrary, for finite quark masses, monomers are present.
%

\section{Chiral and Nuclear Transition}

\subsection{Symmetries and phase diagram}
%
The chiral symmetry at strong coupling is
$\chi'(x)=e^{i\alpha\epsilon(x)}\chi(x)$, where $\epsilon(x) = (-1)^{\sum_{\mu}
  x_{\mu}}$. 
It is spontaneously broken ($\langle \overline{\chi} \chi \rangle \neq 0$) at
low temperatures and densities.
At high temperatures and densities, the chiral symmetry is restored ($\langle
    \overline{\chi} \chi \rangle = 0$).
Between these two phases, there is a 2nd order phase transition line with
$O(2)$ critical exponents at small chemical potential ($\mu<
    \mu_{\text{tric}}$) and a 1st order line $\mu_{\text{1st}}(T) >
\mu_{\text{tric}}$.
The tricritical point (TCP) is located between the 2nd and 1st order lines
point. 
On the other hand, nuclear transition between vacuum phase and nuclear matter
phase does not have the 2nd order line. 
They have the 2nd order CEP that is similar to the CEP of a liquid gas
transition, and the 1st order line is located below the CEP. 
The 1st order line $\mu_{\text{1st}}(T)$ separates the hadronic phase where the
baryon density $\langle n_B \rangle=0$ from the nuclear matter phase. At $T=0$
and above $\mu > \mu_{\text{1st}}$, where $\langle n_B \rangle=1$, Pauli
saturation occurrs:
Due to the finite lattice spacing, the baryons form a crystal in this nuclear
matter phase, i.~e.~every lattice site is filled by a baryon.
Because of the Pauli principle, the mesons and the baryons can not intersect
with each other. 
Hence, in the nuclear matter phase, the chiral condensate $\langle
\overline{\chi} \chi \rangle$ vanishes.
On the contrary, in the hadronic phase, the baryons are rare and the mesons are
common.
%

\subsection{\label{sec:observables}Observables}
%
Our observables for the chiral transition are the chiral condensate $\langle
\overline{\chi} \chi \rangle$ and the chiral susceptibility $\chi_{ch}$.
\begin{align}
\langle \overline{\chi} \chi \rangle = \frac{1}{2m_qV}\langle N_M \rangle\,, 
\quad \chi_{ch} = \frac{1}{V} \frac{\partial^2}{\partial(2m_q)^2}\log{Z} 
= \frac{1}{(2m_q)^2 V} \big( \langle N^2_M \rangle -  \langle N_M \rangle^2 - \langle N_M \rangle \big)\,,
\end{align}
where $N_M \equiv \sum_x n_x$.
In the chiral limit, $\langle \overline{\chi} \chi \rangle = 0$ because
$N_M=0$.
For the nuclear transition, we measure the baryon density $\langle n_B \rangle$
and the baryon susceptibility $\chi_B$, which are given by the winding numbers
$r_{\ell} \in \mathbb{Z}$.
\begin{align}
\langle n_B \rangle = \frac{1}{V_s N_t} \frac{\partial}{\partial(3a_t\mu)}\log{Z} = \frac{1}{V_s} \langle \sum_{\ell} r_{\ell} \rangle\,,
\quad \chi_B = \frac{N_t}{V_s} \big( \langle (\sum_{\ell} r_{\ell})^2 \rangle - \langle \sum_{\ell} r_{\ell} \rangle^2 \big)
\end{align}
The general reweighting method with the average sign is applied in our study.
\begin{align}
\label{eq:sign}
\langle O \rangle = \frac{\langle O \sigma \rangle}{\langle \sigma \rangle}\,, \quad \langle \sigma \rangle = \exp(-L^3 N_t \Delta f)\,,
\end{align}
where $\Delta f$ is the difference between full and sign-quenched free energy
density. 
%

%
\subsection{Anisotropy and finite temperature}
%
We introduce the anisotropy $\gamma$ in the Dirac couplings in order to vary
the temperature in the strong coupling limit $\beta=0$, where $a$ does not
vary, but $a_t$ does vary in Eq.~\eqref{eq:L}.
The ratio of the lattice spacing in spatial and temporal direction can be
written as a general function $\frac{a}{a_t}=\xi(\gamma)$.
Mean field theory of Eq.~\eqref{eq:dual} suggests~\cite{Faldt:1985ec}  that
$\xi'(\gamma)=\gamma^2$, it is an $N_t$-independent choice.
$\xi(\gamma)$ is obtained from non-perturbative calculation~\cite{Helvio}.
Hence, we use the following notations to distinguish $\xi(\gamma)$ and
$\xi'(\gamma)$.
\begin{align}
a\mu'=a_t\mu\xi'(\gamma)\,, \quad aT'=\frac{\xi'(\gamma)}{N_t}\,, \qquad
a\mu=a_t\mu\xi(\gamma)\,, \quad aT=\frac{\xi(\gamma)}{N_t}
\end{align}
%

\section{Results}
\subsection{Average sign}
%
\begin{figure}[!hpbt]
\center
\includegraphics[width=0.9\textwidth, trim=0 5 0 20, clip]{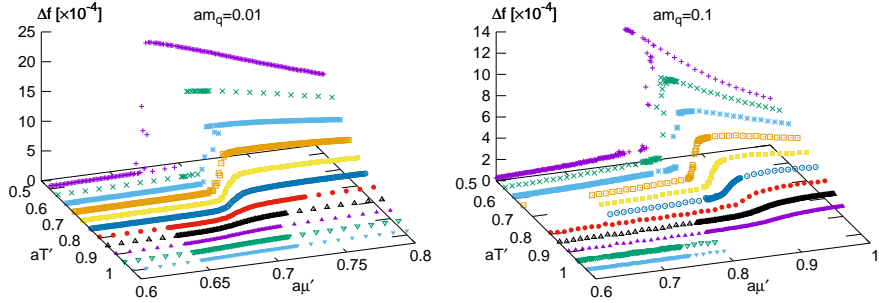}
\caption{
\label{fig:sign}
The average sign in the range of $0.5 < aT' < 1$, and $0.6 < a\mu' < 0.8$ for
$am_q=0.01, 0.1$.
Here, $\Delta f = -\frac{1}{V}\log{\langle \text{sign} \rangle}$
}
\end{figure}
In order to show the mildness of the sign problem in the dual the
representation, we plot $\Delta f$ in Fig.~\ref{fig:sign} which is defined in
Eq.~\eqref{eq:sign}.
In the area of $0.5 < aT' < 1$, and $0.6 < a\mu' < 0.8$, where our simulation
is done, the average sign is quite small for various quark masses $am_q=0.01,
0.1$.
%

\subsection{Finite size scaling at finite density}
%
We use finite size scaling to the chiral and nuclear susceptibilities to find
the temperature of CEP.
The finite size scaling is carried out using the following critical exponents.
In the chiral limit ($m_q=0$), the $O(2)$ exponents with $L^{\rho( \equiv
    \gamma/\nu)}$, $\gamma=1.3177$ and $\nu=0.67155$ are used for the 2nd order
line in chiral transition and CEP in nuclear transition.
For the crossover region in the nuclear transition, $\rho=1$ is applied.
We use the exponents with $\gamma=1$ and $\nu=0.5$ at the TCP for the chiral
transition.
For the finite quark masses, we apply the $Z(2)$ exponents with $\gamma=1.237$
and $\nu=0.613$ at the CEP for both chiral and nuclear transitions.
$\rho=3$ is applied for the first order lines. 
\\
We scan the parameter space along the $\mu$-direction for various temperatures
in the range of $0.5 \le aT \le 1.0$ with the step size $0.05$ to find the CEP
and phase boundary.
We analyse a peak of the chiral and baryon susceptibilities. 
In the chiral limit, the chiral susceptibility does not have a peak. 
Hence, we use the baryon susceptibility to find the TCP temperature because the
location of the TCP in the chiral transition is same as that of the CEP for the
nuclear transition in the strong coupling limit ($\beta = 0$).
For finite quark masses, the chiral susceptibility has a peak. 
So, we obtain the CEPs and phase boundaries separately from the chiral and
nuclear transitions.
We use the standard finite scaling method to find the temperature of the CEP.
We compare the peak heights of the different lattice volumes, they are rescaled
by the CEP exponents.
By the standard method, the peak heights of the different lattice volumes
become equal at the temperature ($aT'_E$) of the CEP. 
\begin{figure}[!htbp]
\center
\subfigure[Fit to the Breit-Wigner with polynomial]{
\label{fig:BW}
\includegraphics[width=0.4\textwidth]{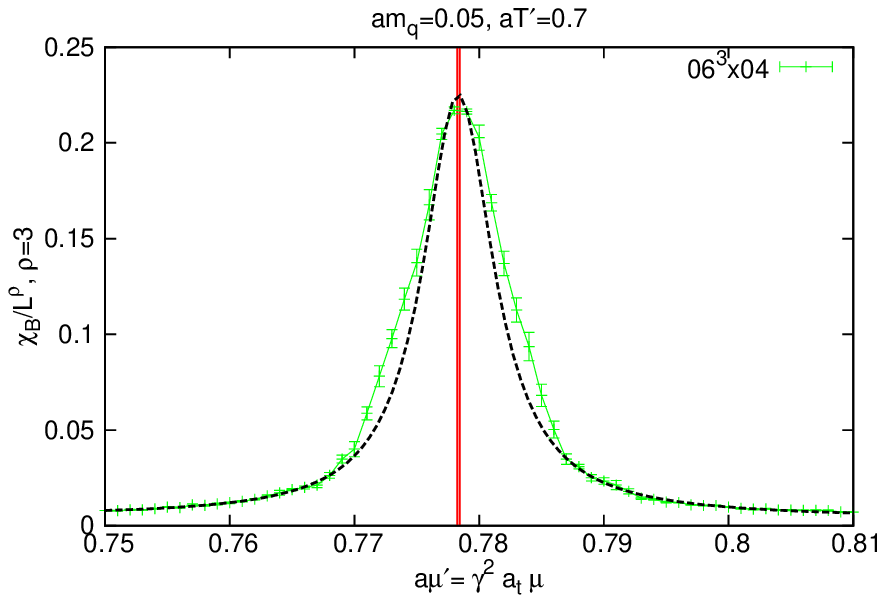}
}
\subfigure[Extrapolation to Thermodynamic limit]{
\label{fig:thermo}
\includegraphics[width=0.4\textwidth]{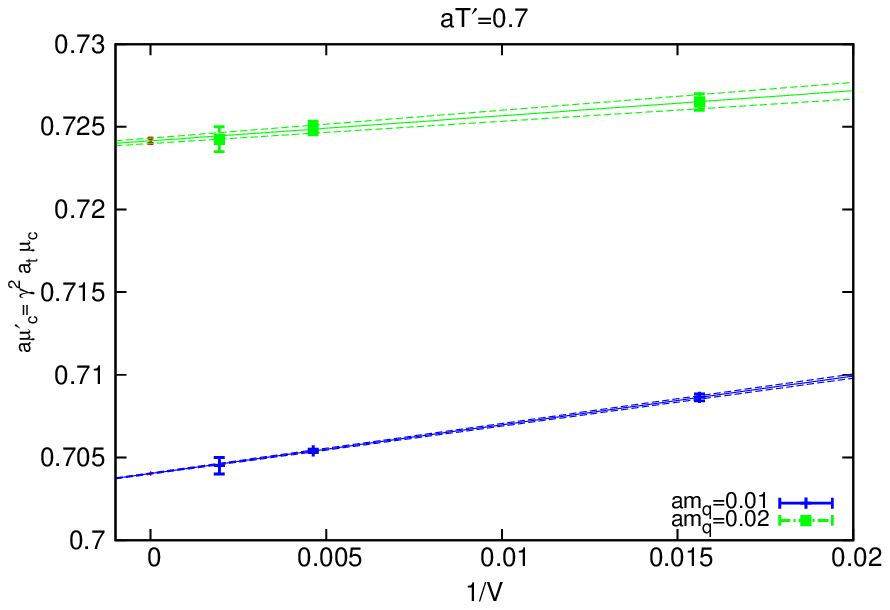}
}
\caption{
The left panel shows the Breit-Wigner fit to find the peak position.  
The right panel shows the extrapolation to the thermodynamic limit for various
quark masses in fixed temperature $aT'=0.7$.
}
\end{figure}
To obtain the critical chemical potential $a_t\mu'_c$, we analyse the peak
position of the chiral and baryon susceptibilities.
For the chiral transition in the chiral limit, we obtain the critical chemical
potential $a_t\mu'_c$ from the crossing points between the different lattice
volumes because the chiral susceptibility does not have a peak.
For other cases, we obtain the peak position using the following way.
First, we fit a peak of the susceptibility using the Breit-Wigner function with
polynomial to find the peak position. 
The red lines in Fig.~\ref{fig:BW} are the peak position with fitting errors.
After we get the $a_t\mu'_c$, we do the extrapolation to the thermodynamic
limit to eliminate the volume dependency. 
The results of extrapolations have very linear behavior with respect to $1/V$
as shown in Fig.~\ref{fig:thermo}.
\\

\begin{figure}[!htbp]
\center
\includegraphics[width=1\textwidth]{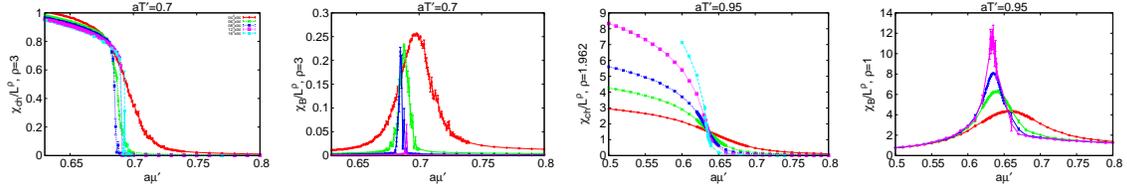}
\caption{
\label{fig:sus_chiral}
The chiral and baryon susceptibilities in the chiral limit.  
In the case of chiral susceptibility, we apply the finite size scaling with
$\rho=1.962$ for the 2nd order in the third panel, and $\rho=3$ for the 1st
order transition in the first panel.
For the baryon susceptibility, we apply $\rho=1$ for crossover transition in
the forth panel, and $\rho=3$ for the 1st order transition in the second panel.
}
\end{figure}
\begin{figure}[htbp]
\center
\subfigure[Chiral transition]{
\label{fig:chiral_finite}
\includegraphics[width=0.47\textwidth]{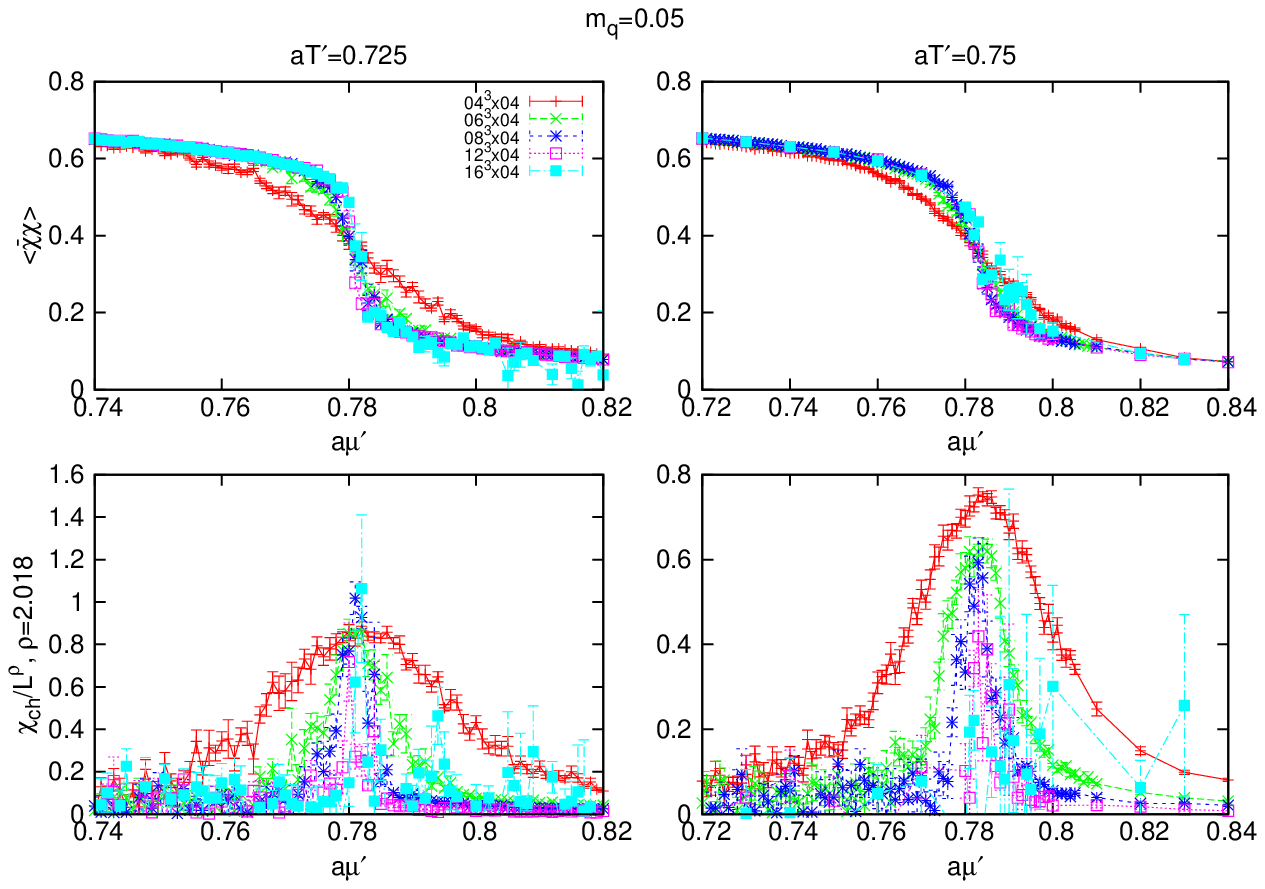}
}
\subfigure[Nuclear transition]{
\label{fig:nucl_finite}
\includegraphics[width=0.47\textwidth]{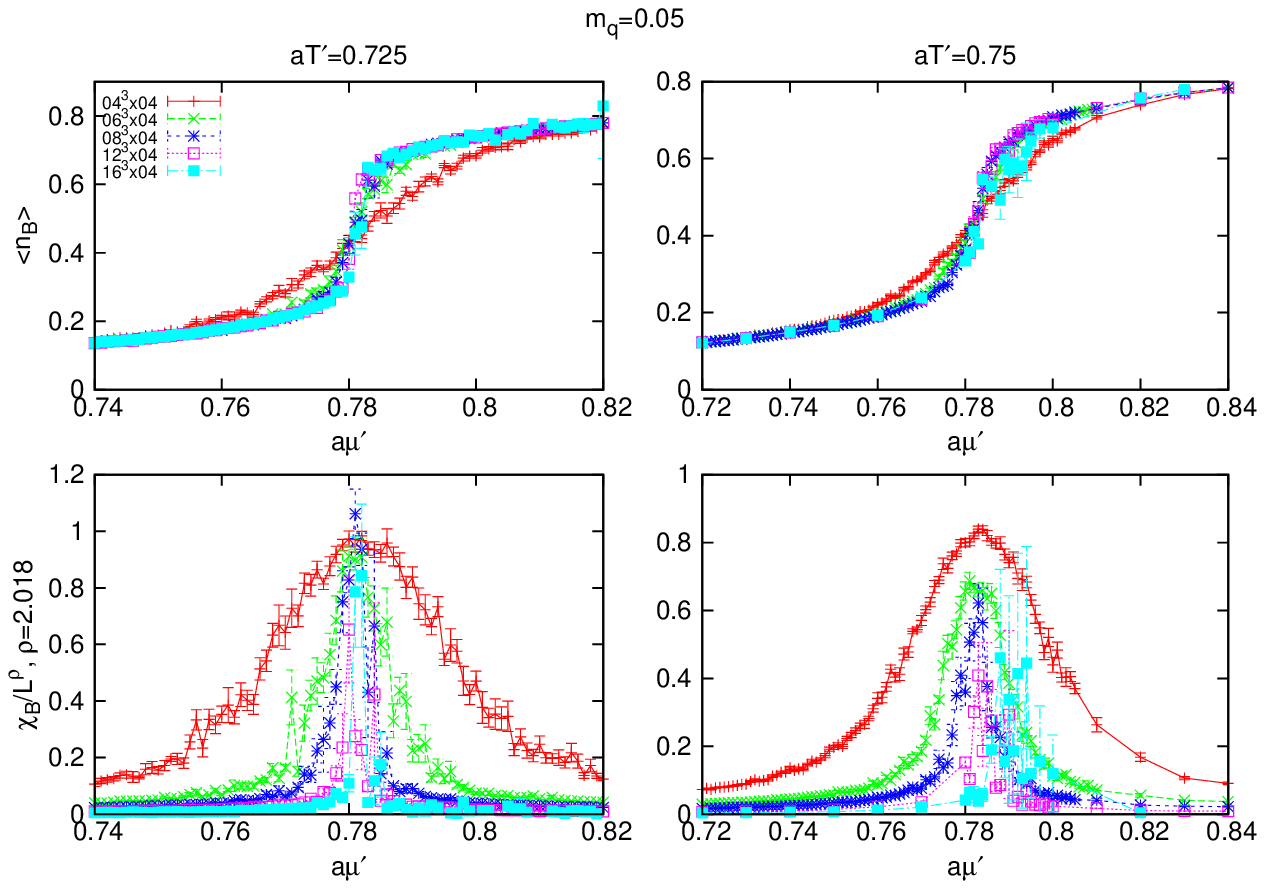}
}
\caption{
The figure (a) shows the chiral condensate and the chiral susceptibility for
finite quark mass $m_q=0.05$ near the critical end point temperature.  
The data is more noisy than those of nuclear transition due to the monomer
fluctuations.
The figure (b) shows the baryon density and the baryon susceptibility for
finite quark mass $m_q=0.05$ near the critical end point temperature.
}
\end{figure}

First, we address the results in the chiral limit comparing the 1st order lines
from the nuclear and chiral transition.
We plot the chiral and nuclear susceptibilities in Fig.~\ref{fig:sus_chiral}.
For the chiral transition at $aT'=0.95$, we use the $O(2)$ exponents for 2nd
line.
But the case of nuclear transition at $aT'=0.95$, crossover scaling $\rho=1$ is
applied.
We apply finite size scaling with $\rho=3$ at $aT'=0.7$ for both chiral and
nuclear transitions because they are belonged in the temperature of 1st order
transition.
If we turn on the quark mass, the chiral susceptibility has a peak.
We plot the chiral condensate and susceptibility for finite quark mass in
Fig.~\ref{fig:chiral_finite}.
The baryon density and susceptibility for finite quark mass are plotted in
Fig.~\ref{fig:nucl_finite}.
For the lower panels in Fig.~\ref{fig:chiral_finite} and
Fig.~\ref{fig:nucl_finite}, the $Z(2)$ exponents are applied. 
Then, the order of peak heights at $aT'=0.725$ and those at $aT'=0.75$ are
opposite.
Hence, we find that the temperature of the CEP is located between $0.725 < aT'
< 0.75$.
The phase boundaries for finite quark masses are obtained from the peak
analysis explained above.
In the Table~\ref{tab:CEP}, our results of CEPs are agree with the previous
Monte Carlo results for various quark masses.
We also compare the results to the Mean Field results in Table~\ref{tab:CEP}.
\begin{table}[hbtp]
\center
\small
\begin{tabular}{c  l  l  l l}
\hline
\hline
$am_q$ & Previous MC($aT'_E$, $a\mu'_E$) & MeanField($aT'_E$, $a\mu'_E$) & Ours($aT'_E$, $a\mu'_E$) &  Ours($aT_E$, $a\mu_E$)\\
\hline
0.00   & 0.94(7), $0.64^{+0.02}_{-0.04}$ & 0.866, 0.577 & 0.83(3), 0.6671(2)  & 0.69(3), 0.5563(2) \\
0.01   & 0.77(3), 0.70(2)                & 0.764, 0.583 & 0.78(3), 0.7005(5)  & 0.66(3), 0.5906(4) \\
0.02   &     N/A                         & N/A          & 0.75(3), 0.7234(14) & 0.64(3), 0.6137(12)\\
0.05   &     N/A                         & 0.690, 0.617 & 0.73(3), 0.7808(5)  & 0.62(3), 0.6653(4) \\
0.10   & 0.69(1), 0.86(1)                & 0.646, 0.653 & 0.70(3), 0.8606(10) & 0.60(3), 0.7386(9) \\
\hline
\hline
\end{tabular}
\caption{
\label{tab:CEP}
We compare our results of CEPs for various quark masses to the previous Monte
Carlo results~\cite{Fromm} and Mean Field results~\cite{Nishida:2003fb}.
The forth column shows the results when we apply the correct anisotropy
$\xi(\gamma)$.
}
\end{table}

\subsection{Phase diagram for finite quark masses}
\begin{figure}
\includegraphics[width=1.0\textwidth]{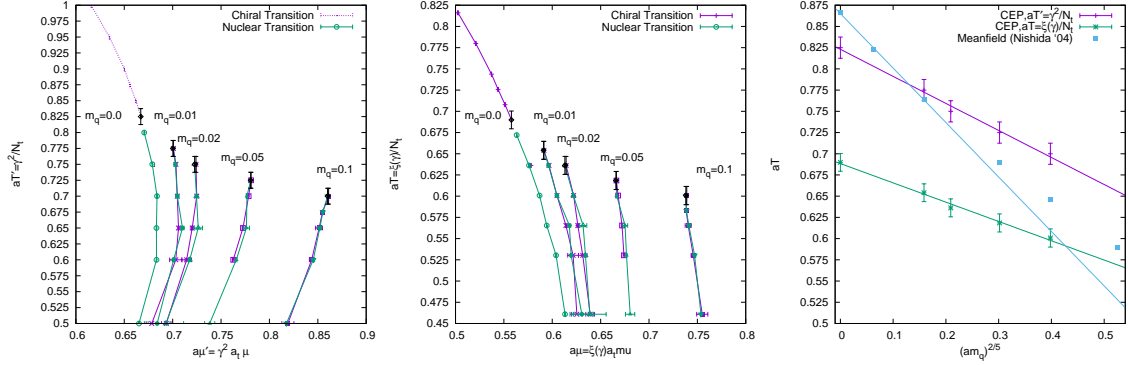}
\caption{
\label{fig:transition}
The first and second panels are the phase boundary of chiral and nuclear
transition for finite quark masses. 
In the first panel, we apply the $\xi'(\gamma)=\gamma^2$ for anisotropy.  
They have back-bending at the low temperature region.
In the second panel, we apply $\xi(\gamma)$ obtained from non-perturbative
calculation. 
After we apply the correct anisotropy, the back-bending has disappeared.
The third panel is the trajectory of critical end points (from $\xi(\gamma)$
and $\xi'(\gamma)$) and those of mean-field theory.
}
\end{figure}

Finally, we obtain the phase diagram for finite quark masses.
We plot the phase diagrams applied $\xi'(\gamma)=\gamma^2$ and $\xi(\gamma)$ in
the first and second panels in Fig.~\ref{fig:transition}. 
When we apply the $\xi'(\gamma)$, they have back-bending in the low temperature
region.
This is because for $\gamma < 1$ spatial dimers are favored, which results in
an unphysical phase boundary.
However, the back-bending disappears when the correct non-perturbative result
$\xi(\gamma)$ is applied.
The third panel shows the trajectory of CEPs and those of mean-field
theory~\cite{Nishida:2003fb}.
The x-axis $(am_q)^{2/5}$ in this panel is suggested by tricritical scaling.
Due to the correct non-perturbative anisotropy $\xi(\gamma)$, the mismatch with
mean-field theory has been enlarged.
Just as $aT_c$ differs between Monte Carlo and mean-field theory, also the
slope in $(am_q)^{2/5}$ differs. 
\section{Conclusion}
We obtain the phase boundary and critical end points for various quark masses
using Monte Carlo simulation in the dual representation.
We extend the 1st order phase boundary to the lower temperature than the
previous Monte Carlo results.
As expected, both the nuclear and chiral 1st order transitions are on top also
for $m_q > 0$.
By applying the non-perturbative results for $\frac{a}{a_t} \equiv
\xi(\gamma)$, we confirm the disappearance of back-bending for all quark
masses. 
\acknowledgments
Calculations leading to the results presented here were performed on resources
provided by the Paderborn Center for Parallel Computing.
We would like to thank Philippe de Forcrand and Helvio Vairinhos for helpful
discussions.
\bibliography{refs}

\providecommand{\href}[2]{#2}\begingroup\raggedright\begin{thebibliography}{1}

\bibitem{Rossi:1984cv}
P.~Rossi and U.~Wolff, {\it {Lattice {QCD} With Fermions at Strong Coupling: A
  Dimer System}},  {\em Nucl. Phys.} {\bf B248} (1984) 105--122.

\bibitem{Adams:2003cca}
D.~H. Adams and S.~Chandrasekharan, {\it {Chiral limit of strongly coupled
  lattice gauge theories}},  {\em Nucl. Phys.} {\bf B662} (2003) 220--246,
  [\href{http://xxx.lanl.gov/abs/hep-lat/0303003}{{\tt hep-lat/0303003}}].

\bibitem{Fromm}
M.~Fromm, {\it {Lattice QCD at string coupling: thermodynamics and nuclear
  physics}},  {\em Thesis} (2010).

\bibitem{deForcrand:2009dh}
P.~de~Forcrand and M.~Fromm, {\it {Nuclear Physics from lattice QCD at strong
  coupling}},  {\em Phys. Rev. Lett.} {\bf 104} (2010) 112005,
  [\href{http://xxx.lanl.gov/abs/0907.1915}{{\tt 0907.1915}}].

\bibitem{Unger:2011in}
W.~Unger and P.~de~Forcrand, {\it {Continuous Time Monte Carlo for Lattice QCD
  in the Strong Coupling Limit}},  {\em PoS} {\bf LATTICE2011} (2011) 218,
  [\href{http://xxx.lanl.gov/abs/1111.1434}{{\tt 1111.1434}}].

\bibitem{Faldt:1985ec}
G.~Faldt and B.~Petersson, {\it {Strong Coupling Expansion of Lattice Gauge
  Theories at Finite Temperature}},  {\em Nucl. Phys.} {\bf B265} (1986)
  197--222.

\bibitem{Helvio}
P.~de~Forcrand, P.~Romatschke, W.~Unger, and H.~Vairinhos, {\it {Thermodynamics
  of strongly-coupled lattice QCD in the chiral limit}},  {\em PoS} {\bf
  LATTICE2016} (2016) 086.

\bibitem{Nishida:2003fb}
Y.~Nishida, {\it {Phase structures of strong coupling lattice QCD with finite
  baryon and isospin density}},  {\em Phys. Rev.} {\bf D69} (2004) 094501,
  [\href{http://xxx.lanl.gov/abs/hep-ph/0312371}{{\tt hep-ph/0312371}}].

\end{thebibliography}\endgroup
\end{document}